# Chalcogenide optical parametric oscillator

Raja Ahmad[1,*] and Martin Rochette[1]

[1] *Department of Electrical and Computer Engineering, McGill University, Montreal (QC), Canada, H3A 2A7.*
[*]*raja.ahmad@mail.mcgill.ca*

**Abstract:** We demonstrate the first optical parametric oscillator (OPO) based on chalcogenide glass. The parametric gain medium is an $As_2Se_3$ chalcogenide microwire coated with a layer of polymer. The doubly-resonant OPO oscillates simultaneously at a Stokes and an anti Stokes wavelength shift of >50 nm from the pump wavelength that lies at $\lambda_P$ = 1552 nm. The oscillator has a peak power threshold of 21.6 dBm and a conversion efficiency of >19 %. This OPO experiment provides an additional application of the chalcogenide microwire technology; and considering the transparency of $As_2Se_3$ glass extending far in the mid-infrared (mid-IR) wavelengths, the device holds promise for realizing mid-IR OPOs utilizing existing optical sources in the telecommunications wavelength region.

## 1. Introduction

When the optical parametric gain provided by a pump laser is trapped in a resonant cavity, the resulting laser operating at a wavelength that differs from the pump laser is termed as an optical parametric oscillator (OPO). OPOs are attractive owing to their tunability over a broad frequency range that cannot be easily covered with conventional laser sources. OPOs have been realized in different nonlinear materials, based on both the second order $\chi^{(2)}$ and the third order $\chi^{(3)}$ susceptibility, and in a variety of device structures and configurations. The OPOs based on $\chi^{(2)}$ nonlinear materials including $LiNbO_3$, KTP, $BiB_3O_6$, MgO:sPPLT [1-5] are bulky, susceptible to misalignment, and can only oscillate at wavelengths longer than that of the pump laser. On the other hand, $\chi^{(3)}$ based OPOs have been realized in conventional and microstructured optical fibers (MOFs), planar waveguides and microresonators made from various materials including silica and silicon [6-15]. However, the high power requirements and long nonlinear interaction lengths required for conventional silica fiber based OPOs render them impractical. The OPOs based on integrated or microresonator devices on the other hand, are attractive in terms of compactness but are not compatible with the existing fiber technology. Furthermore the microresonator based OPOs suffer from longer response times, from hundreds of picoseconds to hundreds of nanoseconds, and the silicon based OPOs suffer from large two-photon absorption and free-carrier generation as well. The MOF-OPOs on one hand, are attractive for their fiber compatibility and their well controlled dispersion profile – that is critical for nonlinear parametric processes – but have only been realized using silica glass that itself suffers from relatively weak nonlinearity.

Among the common third order nonlinear optical materials, arsenic triselenide ($As_2Se_3$) chalcogenide glass boasts one of the highest nonlinear refractive index coefficient in a glass $n_2 = 2.3 \times 10^{-13}$ $cm^2/W$ [16], that is up to 1000 × that of silica, 20 × that of $Bi_2O_3$, 4 × that of $As_2S_3$, and 3 × that of Si [16-18]. Despite the large value of $n_2$ in $As_2Se_3$, the material exhibits a large normal material chromatic dispersion in the 1550 nm wavelength band, where there is an abundance of lasers that could be utilized for OPO pumping. This chromatic dispersion prevents efficient parametric gain close to the pump wavelength, thereby rendering it difficult to realize OPOs from $As_2Se_3$ in bulk or in an optical fiber format. One way to overcome this limitation is to use $As_2Se_3$ microwires for which the anomalous waveguide dispersion overcomes the normal material dispersion. Such microwires in addition, exhibit large values of waveguide nonlinear coefficient $\gamma$ $(= n_2\omega_P/cA_{eff}$, $\omega_P$ being the pump angular frequency, $c$ being the speed of light and $A_{eff}$ being the effective mode area in the microwire) which lowers the required power threshold for nonlinear applications including the realization of OPO. The highest $\gamma$ value reported in such microwires is more than 5 orders of magnitude larger than that of silica fibers [19]. However, to our knowledge, no chalcogenide OPO in any configuration has yet been reported in the literature.

In this paper, we report an OPO based on an $As_2Se_3$ microwire coated with poly-methyl meth-acrylate (PMMA) cladding. The PMMA cladding in addition to providing physical strength to the thin microwire [20], serves to optimize phase-matching conditions towards efficient and broadband parametric gain [21]. The OPO oscillates simultaneously at two wavelengths: at a Stokes and anti-Stokes wavelength shifts of +53 nm and -50 nm respectively from the pump laser. The large nonlinearity, the reduced chromatic dispersion from a PMMA cladding and the long effective length (due to the low absorption loss α < 1 dB/m ) of $As_2Se_3$ hybrid microwires allows the OPO to oscillate at a low peak pump power threshold of 21.6 dBm (pulse energy = 3.15 pJ) and with a total conversion efficiency of > 19%.

## 2. Experimental Results and Discussion

The hybrid $As_2Se_3$-PMMA microwire is prepared following the procedure detailed in Ref. [19,21]. The microwire has an $As_2Se_3$ core diameter of 1.01 µm and is 10 cm long, with a total insertion loss of 5 dB. Fig. 1 shows the experimental setup for the OPO operation. A mode-locked laser emitting pulses of full-width at half maximum (FWHM) duration of ~450 fs, with a repetition rate of 20 MHz and spectrally centered at around λ =1,551 nm, is used as the pump laser. The pump is filtered spectrally with a ~0.25 nm bandpass filter (BPF) centered at 1552 nm to lengthen pulses to a FWHM duration ~ 22ps [20]. The resulting pump pulses are then delivered to the microwire via the 10 % output port of a 90/10 fiber coupler (FC). This results in the optical parametric amplification on both sides of the pump wavelength in an almost symmetric manner, determined precisely by the dispersion profile of the microwire. In order to realize OPO, the parametric gain obtained is fed-back to the microwire via the second input port of the FC, thus completing the laser cavity. The residual pump pulse is filtered out from the fed-back signal using a fiber Bragg grating resonant at the pump wavelength combined with an optical fiber circulator (CIR). This avoids any unfavourable interference of the incoming pump pulse with the preceding residual one. A fiber-coupled tunable optical delay line (ODL) is introduced for a precise control of the cavity length so that the amplified Stokes and anti-Stokes signals that resonate in the loop cavity are precisely synchronized with the incoming pump pulse. It is noted that the dispersive walk-off effect can be ignored since the walk-off length of ~ 2.2 m [22], resulting from the wavelength separation between the pump and the Stokes/anti-Stokes signals is considerably longer than the length of the parametric gain medium i.e., the 10 cm long microwire. Finally, a fiber polarization controller is inserted in the cavity to align the polarization state of the two amplified signals with that of the incoming pump pulse. The operation of the OPO is observed on an optical spectrum analyzer (OSA) connected to the 90 % output port of the fiber coupler.

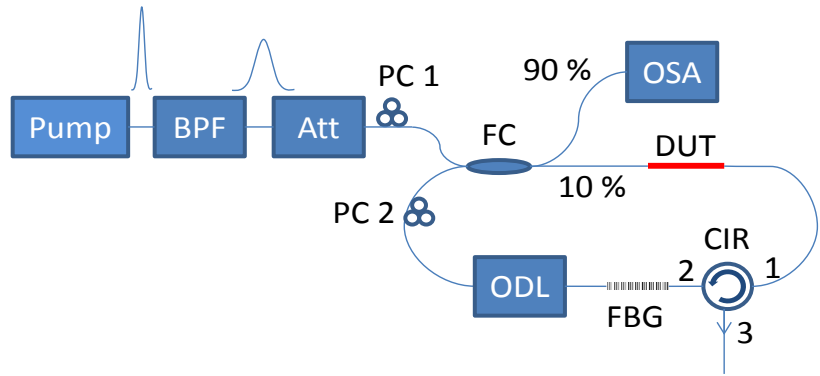

**Fig. 1. Experimental setup for the OPO operation.** BPF: band-pass filter; Att: optical attenuator; PC: fiber polarization controller; FC: fiber coupler; OSA: optical spectrum analyzer; DUT: device-under test; CIR: optical circulator; FBG: silica fiber bragg grating; ODL: optical (tunable) delay line.

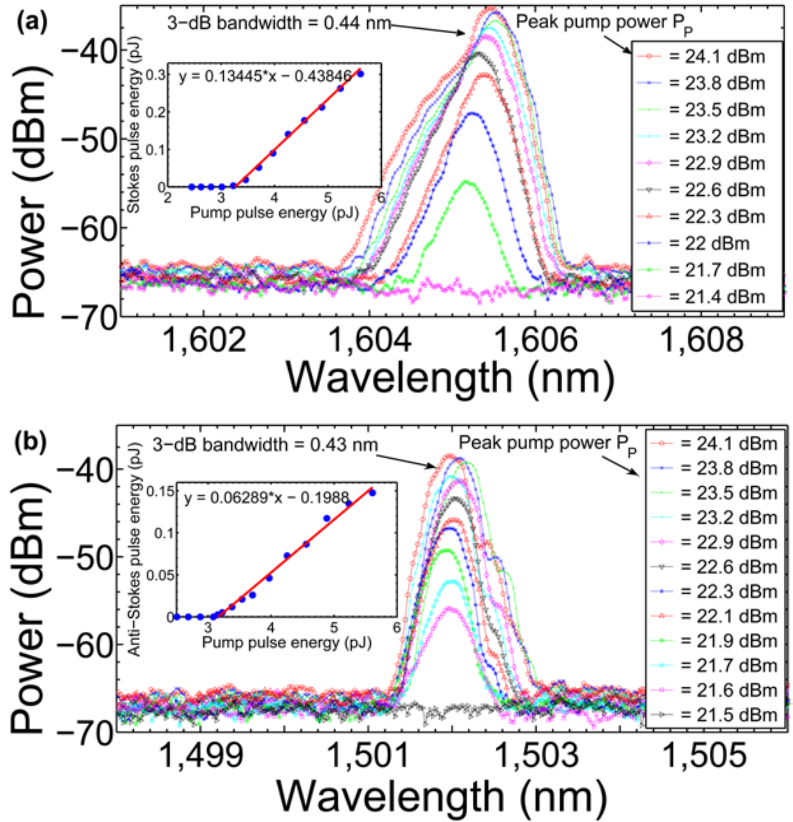

**Fig. 2. Output spectra of the (a) Stokes and (b) anti-Stokes OPO signals** for the increasing values of input peak pump power. (Inset) The pulse energy in the Stokes and the anti-Stokes output signals are plotted against the input pump pulse energies and are included as inset in (a) and (b) respectively.

When the total single-pass parametric gain exceeds the round-trip cavity loss of 8 dB, the OPO oscillates at Stokes and anti-Stokes wavelengths of 1,605 nm and 1502 nm respectively. This represents the conversion of a C-band pump laser to L-band (Stokes) and S-band (anti-Stokes) OPOs. Figs. 2 (a) and (b) show the spectra of the output Stokes and anti-Stokes OPO signals with increasing peak pump power. The spectral FWHMs of the two OPO outputs are 0.44 nm and 0.43 nm for the Stokes and the anti-Stokes signals respectively. The Stokes signal carries 3.3 dB more energy than the anti-Stokes one. This can be explained from the additional Raman gain at the Stokes wavelength that almost coincides with the Raman shift wavelength for the $As_2Se_3$ microwire [21]. Both the Stokes and the anti-Stokes OPO signals have a threshold peak pump power of ~21.6 dBm, corresponding to pulse energy of 3.15 pJ. The slope efficiency of the Stokes OPO exceeds 13 %, with the Stokes output pulses carrying > 0.3 pJ of energy. The slope efficiency of the anti-Stokes OPO is >6 % corresponding to a pulse energy ~ 0.15 pJ, providing a total internal conversion efficiency of > 19 %. This represents high conversion efficiency OPO considering its compactness and the low-power operation.

Fig. 3 shows the output spectra of the OPO as recorded on the OSA. The spectra show the simultaneous oscillation of the doubly-resonant OPO at the Stokes and anti-Stokes wavelengths. After setting up the OPO, we study the effect of adding a delay or advance on the oscillating Stokes and anti-Stokes signals which compromises their precise synchronization with the pump pulse. The OPO continues to operate for a wide temporal detuning of approximately ± 5 ps beyond which there is no output signal. This temporal detuning allows OPO Stokes and anti-Stokes wavelengths tuning by up to 8 nm. The wavelength tuning results from a natural minimization of the group velocity mismatch between the pump wavelength and the Stokes/anti-Stokes oscillating wavelengths. In order to compare the experimental results with theory, the single-pass parametric gain in the microwire of the same dimensions as used in the experiment is calculated for a 24.1 dBm peak pump power. The calculated parametric gain profile is included in the Fig. 3. It is noted that the oscillating wavelengths precisely match the wavelengths where the parametric gain is maximum in agreement with theory.

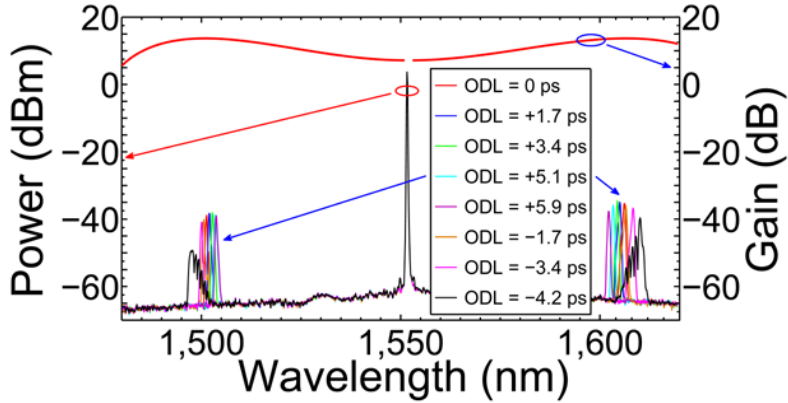

**Fig. 3. Output of the OPO as observed on an OSA** for various delay values on the oscillating Stokes and anti-Stokes signals. The calculated parametric gain spectrum under the experimental conditions is also included.

## 3. Conclusion

In conclusion, we have demonstrated the first optical parametric oscillator in chalcogenide glass. The parametric gain medium is a 1.01 µm thick and 10 cm long $As_2Se_3$-PMMA hybrid microwire. The OPO oscillates simultaneously at Stokes and anti-Stokes wavelengths in the L- and the S- telecommunications frequency bands respectively, with the pump lying in the C-band. The OPO has a total internal conversion efficiency of ~ 19 %, with a threshold peak pump power of 21.6 dBm. The wavelength conversion bandwidth of this device can be further extended to the mid-infrared wavelengths region by pumping with the existing C-band laser but by adjusting the wire diameter to have a net, small value of normal dispersion at the pump wavelength.

## Acknowledgments

The authors would like to thank Coractive High-Tech inc. for providing the Chalcogenide fiber used in the experiments. This work was financially supported by FQRNT (Le Fonds Quebecois de la Recherche sur la Nature et les Technologies) and the Natural Sciences and Engineering Research Council of Canada (NSERC).